\documentclass[onecolumn,12pt,draftclsnofoot]{IEEEtran}
\usepackage{cite}
\usepackage{amsmath,amssymb,amsfonts}
\usepackage{algorithmic}

\usepackage{xcolor}
\usepackage{amssymb}
\usepackage{bbm}
\usepackage{amsmath}
\usepackage{bm}
\usepackage{amssymb,mathtools}
\usepackage{amsmath}
\usepackage{amsbsy}
\usepackage[font=small]{subcaption}
\usepackage{graphicx} 

\usepackage{varioref}
\usepackage{xcolor}

\newcommand*{\field}[1]{\mathbb{#1}}
\def\BibTeX{{\rm B\kern-.05em{\sc i\kern-.025em b}\kern-.08em
    T\kern-.1667em\lower.7ex\hbox{E}\kern-.125emX}}

\newcommand{\fsprob}{p_\mathrm{f}}
\newcommand{\fsprobt}{p}
\newcommand{\N}{N}

\newcommand{\area}{\mathcal{A}}
\newcommand{\crit}{\mathrm{cr}}

\begin{document}

\title{Modeling and Analysis of Wildfire Detection using Wireless Sensor Network with Poisson Deployment}

\author{\IEEEauthorblockN{ Kaushlendra Pandey, Abhishek Gupta}
\thanks{The authors are with the Modern Wireless Networks Group, Electrical Engineering at Indian	Institute of Technology Kanpur, Kanpur (India) 208016 (Email:\{kpandey,gkrabhi\}@iitk.ac.in).}
}

\newcommand{\insertnotationtable}{
 
\begin{table}
\caption {\small Notation Table} \label{tab:title}
\begin{center}
    \begin{tabular}{| l | l |}
    \hline 
    \textbf{Symbol} & \textbf{Definition}  \\ \hline 
    $\Phi$ & Homogeneous PPP  which models locations  of sensor nodes in the network. \\ \hline
    $\lambda$ & Density of wireless sensor network per unit area.\\ \hline
    $r_i$& The (random) sensing radius of $i$th sensor.\\\hline
$x_i$ & The location of $i$th sensor.\\\hline
$\mathsf{S}_i$ & $\mathcal{B}(0,r_i)$\\\hline
    $\xi$ & Combined covered area of all sensors.\\ \hline
 $\fsprobt{(t)}$ & Sensing probability of set $\mathcal{K}$ at time $t$.\\ \hline
  $\mathcal{B}(x,r)$ & Ball of radius $r$ centred at $x$.\\ \hline
     $\oplus$ & Minkowski addition.\\ \hline
      $r_{\mathrm{in}}$ & Fixed sensing range of a sensor node.\\ \hline
       $r_{\mathrm{out}}$ & Maximum sensing range of a sensor node.\\ \hline
       $\area(.)$ & Area of a set (.).\\ \hline
        $\ell(.)$ & Perimeter of a set (.).\\ \hline
         $A_\crit, t_\crit$ & The critical area of fire and critical time to reach it.\\ \hline
          $\fsprob$ & Fire detection probability of fire before it goes critical.\\ \hline       
    \end{tabular}
\end{center}

\end{table}}

\maketitle

\begin{abstract}

In many forest fire incidences, late detection of the fire has lead to severe damages to the forest and human property requiring more resources to gain control over the fire. An early warning and immediate response system can be a promising solution to avoid such massive losses. This paper considers a network consisting of multiple wireless sensors randomly deployed throughout the forest for early prompt detection of fire.  We present a framework to model fire propagation in a forest and analyze the performance of considered wireless sensor network in terms of fire detection probability.  In particular, this paper models sensor deployment as a Poisson point process (PPP) and models the forest fire as a dynamic event which expands with time. We also present various insights to the system including required sensor density and impact of wind velocity on the detection performance. We show that larger wind velocity may not necessarily imply bad sensing performance or the requirement of a denser deployment.

\end{abstract}

\section{Introduction}
\IEEEPARstart{W}{ildfire} is one of the most dangerous natural calamity, which not only hampers the ecosystem and biodiversity of forests, but also results in great loss of human lives and property. It has been estimated that  50\% of the total forest cover in India is affected by occasional fire events with 6\% of the area being at high and frequent risks of wildfire \cite{satendra2014forest}. According to a report by the Department of Science and Technology (Government of India) \cite{sabha2005department}, a forest fire,  in the month of February to April in 2016, smashed nearly 4000 hectares of forest-cover spanning over 13 districts of the state Uttarakhand (India). The report also discussed various causes of forest fire and precautionary steps to control wildfires, for example, construction of watch towers for detecting forest fire, deployment of forest watchers, creation and maintenance of fire-line and use of remote sensing technologies. 
However, there are some practical constraints in implementing the above-mentioned measures such as inadequate infrastructure, heavy cost, and unskilled staff. These constraints may lead to a delayed response in the case of a fire-event. This can result in a multi-fold damage to the forest and human property that could have avoided with an early response.

One way to build an efficient alarm system for early detection of wildfire is with the use of wireless sensor network (WSN) of fire sensors deployed in forests. Wireless sensor network refers to a network of connected wireless sensors and has gained popularity as a cost-effective inexpensive solution to jointly detect an event/events including fire over an area owing to the recent advancements in the sensor technology and wireless communication. 
 
Wireless networks of these deployed nodes can also find a variety of application in the forest including habitat monitoring, wildlife monitoring, humidity variations with the season, and understanding the nature of a particular kind of animal, and its population.

The performance of a WSN can be characterized in terms of its coverage {\em i.e.} the probability that the event is sensed by at least one node of the WSN. The coverage performance of a WSN with sensors having fixed disk sensing range is analyzed in \cite{kasbekar2011lifetime}.
Readers are advised to refer to  \cite{simplot2005energy} for an extensive literature survey discussing the coverage and connectivity analysis of WSNs. Maintenance of coverage and connectivity in a network by activating a minimum number of sensor nodes have been addressed in \cite{zhang2005maintaining}. A study of deployment patterns of sensor node was performed in \cite{yun2010optimal}  to get full coverage and $k$-connectivity under different sensor placement schemes. Energy efficient optimal coverage and full connectivity was studied in  \cite{bai2010optimal,he2013emd}. The deterministic deployment of sensor nodes may not be possible for forest applications where the terrains are not uniform. In such applications, random deployment of sensors can be assumed. Tools from Stochastic geometry provide a tractable framework to study the coverage of random networks including WSN \cite{AndGupDhi16}. Coverage performance of random WSNs was studied in \cite{HaenggiBook, BaccelliBook}. The main limitations of the above-mentioned work is  assumption of the static nature of the event to be sensed.

The events like fire tend to expand their affected area over time.  To analyze the coverage of a fire event using WSN, it is crucial to understand how wildfire grows with time.  Different models were studied in \cite{el2018framework} to understand the dynamics of wildfire propagation. The two primary approaches to model forest fire propagation are raster-based approach and vector based approach \cite{sullivan2009wildland}.  The raster-based approach assumes that the fire propagates from cell to cell under certain propagating conditions. The vector-based based approach assumes that fire grows according to certain geometrical shape which can expand and shift with time. Few of the vector-based fire propagation models were introduced in \cite{coleman1996real}.  In \cite{weber1989analytical}, it was shown that fire propagates according to an expanding circular shape in a homogeneous forest fire with no wind. The elliptical and circular fire front propagation models were discussed in \cite{finney1998farsite}. The work \cite{fendell2018uavs} studied the impact of the wind on the growth of the fire. The coverage performance of a random WSN to sense a time-evolving event has not been studied in the previous work which is the main focus of our paper.

In this paper, we have considered a random wireless fire sensor network (WFSN) deployed in a forest to sense an event of the fire. We develop an analytical framework to model the propagation of wildfire with time in the presence of wind and derive the performance of WFSN in terms of the fire-sensing probability as a function of time passed since the start of the fire. We also characterize the fire detection probability and the critical sensor density to detect a fire before it goes critical/uncontrollable. We also investigate the impact of wind velocity on it and show that a larger wind velocity may not necessarily imply the requirement of a denser deployment.

\begin{figure}
\def\svgwidth{\textwidth}
\centering
  \includegraphics[width=.5\textwidth]{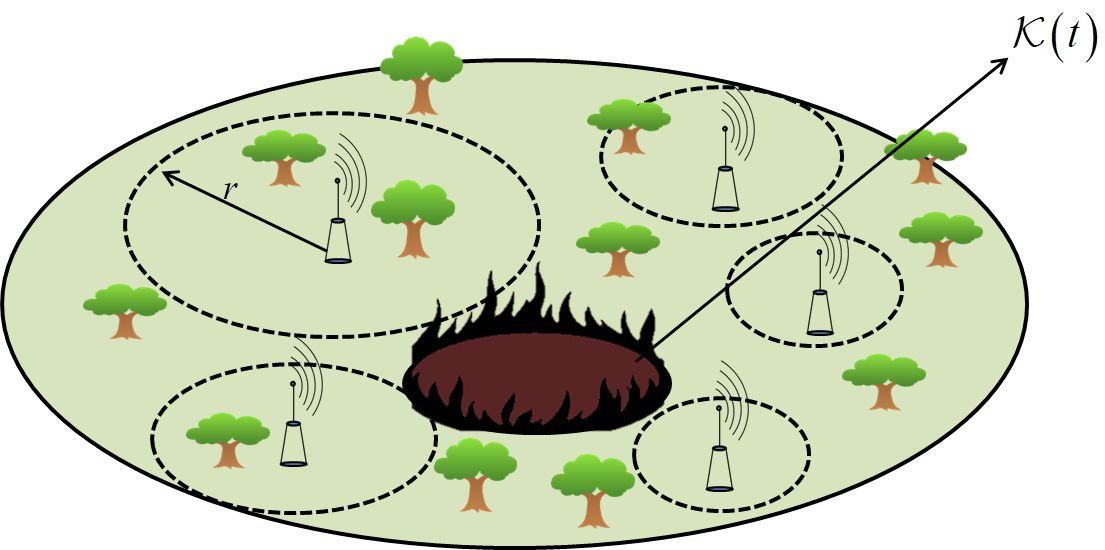}
  \caption{\small Illustration showing a network of wireless fire sensors deployed over the forest. Each sensor has a random sensing range $r$. A fire started at a point grows into the fire-region modeled by the set $\mathcal{K}(t)$ at time $t$.} \label{circularfirefront}
\end{figure}

\section{System model}
 This paper analyzes the early detection of forest fire before it becomes uncontrollable with the help of a randomly deployed network of wireless fire sensors. A list of symbols used in this paper is shown in the table \ref{tab:title}.

 We consider that nodes of the WFSN  are deployed in the 2$d$ space $\mathbb{R}^2$.  Each sensor has a sensing region around it which denotes the region this sensor can sense for fire.  We model the complete network of wireless sensors (locations and sensing regions of sensors) by a boolean model $\Psi$. In this model, the locations of wireless sensors are model as a PPP, and each sensor is assumed to have an identically distributed and independent (iid) sensing zone around it.

  We model the locations of sensors by the PPP $\Phi$ with density $\lambda$, which represents the number of sensors deployed per unit area of the forest. Let $x_{i}$ denote the $i$th wireless sensor location.
    We represent each sensing zone   of \(i\)th sensor as a ball ($\mathcal{B}(x_i,r_{i})$) of radius $r_{i}$ centered at $x_{i}$. Here,  $r_{i}$ is the sensing radius of $i$th sensor and assumed to be a  iid random variable. Let $\mathsf{S}_{i}$ denote $\mathcal{B}(0,r_{i})$. Let us denote \(i\)th sensor by the tuple ($x_{i},r_{i}$) which denotes \(i\)th sensor located at $x_{i}$ with sensing radius $r_{i}$. 
   
  To model $r_i$, we consider the hybrid sensing model which is a combination of disk sensing model and exponential model  \cite{pudasaini2009stochastic}.

 In this hybrid model, the total sensing range of a sensor $x$ is modeled as summation of a fixed sensing range $r_{\mathrm{in}}$ and a random variable $y$:
\begin{equation}
r=r_{\mathrm{in}}+y.
\end{equation}
where $y$ is a truncated exponential random variable between $0$ to $R'$ with probability of density of function (pdf):
\begin{equation}
  f(y) =
  \begin{cases}   
    \frac{e^{-y}}{1-e^{-R'}} & \text{if $0<y\leq R'$}  \\ 
    0       & \text{otherwise.}                          
    \end{cases}       
\end{equation} 
Here $R'=r_{\mathrm{out}}-r_{\mathrm{in}}$ with $r_{\mathrm{out}}$ being the maximum sensing range of sensor. The expected value of $r$ and $r^2$ is given as 
\begin{align}
\mathbb{E}[y]&=\Big(1-\frac{R'e^{-R'}}{1-e^{-R'}}\Big)\nonumber\\
\mathbb{E}[r]&=\frac{1+r_{\mathrm{in}}-(1+r_{\mathrm{out}})e^{-(r_{\mathrm{in}}-r_{\mathrm{out}})}}{1-e^{-(r_{\mathrm{in}}-r_{\mathrm{out}})}}\label{expectedvalue}\\
\mathbb{E}[r^{2}]&=r_{\mathrm{in}}^{2}+2\mathbb{E}[y](1+r_{\mathrm{in}})-\frac{R'^{2}e^{-R'}}{1-e^{-R'}}. \label{finalvariance}
\end{align}

\eqref{expectedvalue} and \eqref{finalvariance} will be later used in calculating Minkowski addition. 

  Now, the total occupied space $\xi$ by the sensors $\Psi$ is a subset of $\mathbb{R}^{2}$ and is represented as
\begin{equation}\label{coveragezone}
\xi=\bigcup_{i \in \field{N}} x_{i} +\mathsf{S}_{i}.
\end{equation}

\insertnotationtable

\section{Modeling of Time-Evolution of Wildfire}

We model the fire-front  including the areas affected by it at time $t$  as a  set $\mathcal{K}(t)$ where $t=0$ denotes the start of the fire.  The dependence on time $t$ represents the dynamic nature of the fire-size. $\mathcal{K}(t)$ can be assumed to be convex \cite{pastor2003mathematical}. Let us define the critical fire-area $A_\mathrm{cr}$ as the area of $\mathcal{K}(t)$ before the fire turns critical (uncontrollable or difficult to manage). The time at which the fire becomes critical is termed as critical time $t_\crit$ 
\begin{align}
t_\crit: \area(\mathcal{K}(t_\crit))=A_\crit.
\end{align}

 In the past literature, various models for fire propagation are used considering the impact of the local environmental condition and the velocity of  wind. In this section, we consider three specific models motivated from the past literature \cite{pastor2003mathematical}.

\subsection{Elliptical Model}
 In \cite{sero2008large,peet1967shape,richards1990elliptical}, the 
authors developed the generalized anisotropic propagation model with non-local radiation term, and proposed the elliptical model as a candidate model. This model is also validated with experimental data and simulation and it was found that fire may not have a steady state rate initially but after some time, it grows with an elliptical geometric shape.

Motivated by these results, we model the dynamic fire ignited on a point as an elliptical shape at any time $t$ with the major axis aligned along the direction of air.   At time $t$, the fire region $\mathcal{K}(t)$ (see Fig. \ref{inkscape}) is given as:
\begin{align}
  \mathcal{K}(t)=\left\{ x,y:
    \begin{tabular}{l}
      $x=t(g+f\cos\phi)$ \\
      $y=t(h\sin\phi)$
\end{tabular},{0\leq\phi\leq2\pi}
\right\}
\end{align}
where $f$, $g$ and $h$ are homogeneous to velocity and are determined by experimental data. It can be seen that the major axis of $\mathcal{K}(t)$ is   $a(t)=ft$, minor axis is $b(t)=ht$ and the center is $(gt,0)$. 

The different values of $a(t)$ and $b(t)$ under the different wind velocity are discussed in the result section of \cite{sero2008large} and it can be concluded that variations in the major and minor axis of the fire region $\mathcal{K}(t)$ can be modeled as:
\begin{align} \label{generalequation}
a(t) &= \alpha t(1 + \frac{{{v_x}}}{V})\\
b(t) &= \alpha t(1 + \frac{{{v_y}}}{V}).
\end{align}

 where $v_{x}$ and $v_y$ are the wind velocities in $x$ and $y$ direction. $V$ is the scaling factor. $\alpha$ is  the firefront velocity in the absence of wind and depends on the other environment conditions and forest density. Without loss of generality that we assume that there is no wind in the $y$-direction ($v_{y}=0$) which gives the following expressions for major and minor axes:
 \begin{equation} \label{majminaxes}
\begin{array}{l}
a(t) \approx \alpha t(1 + \frac{{{v_x}}}{V})\\
b(t) \approx \alpha t.
\end{array}
\end{equation}

The area $\area(.)$ and the perimeter $\ell(.)$ of the set  $\mathcal{K}(t)$ under elliptical model is given as
\begin{align}
\label{areaelliplse}
\area(\mathcal{K}(t))&=\pi.a(t)b(t)\\
\label{perimerterellipse}
\ell(\mathcal{K}(t))&\approx \pi [3(a(t)+b(t))-\sqrt{(3a(t)+b(t))(a(t)+3b(t))}].
\end{align}
\subsection{Circular Model}
In the absence of wind ($v_{x}=v_{y}=0$) the elliptical model converges to a circular model.  Intuitively, we can also see that under the uniform condition such as vegetation and humidity and absence of wind, the fire front will propagate circularly. 

 Fig. \ref{circular1} shows the propagation of  fire ignited at point $O$. As time grows, the radius $r_{\mathcal{K}}(t)$ of the fire affected area  is given by 
\begin{equation}
r_{\mathcal{K}}(t)=\alpha t.
\end{equation}
As discussed earlier, $\alpha$ is the fire velocity in absence of wind which is consistent with the definition.
\subsection{Piriform Model} 
\begin{figure}[ht!]

\begin{subfigure}[ht!]{\linewidth}

 \def\svgwidth{0.5\textwidth} 
 \centering
\includegraphics[width=0.5\textwidth]{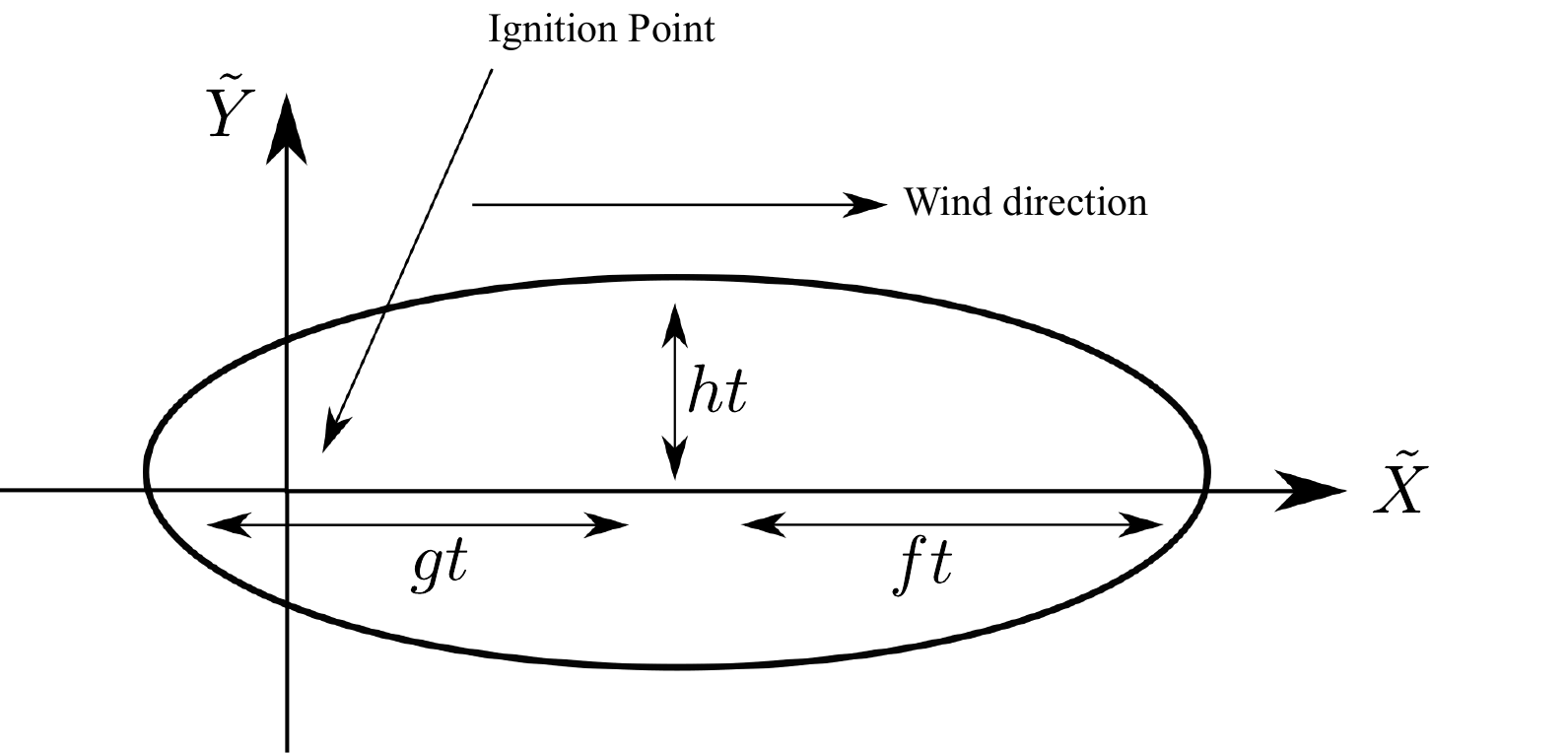}
\caption{}\label{inkscape}

   \end{subfigure} 
   \begin{subfigure}[ht!]{\linewidth}
\def\svgwidth{\textwidth}
\centering
\includegraphics[width=0.3\textwidth]{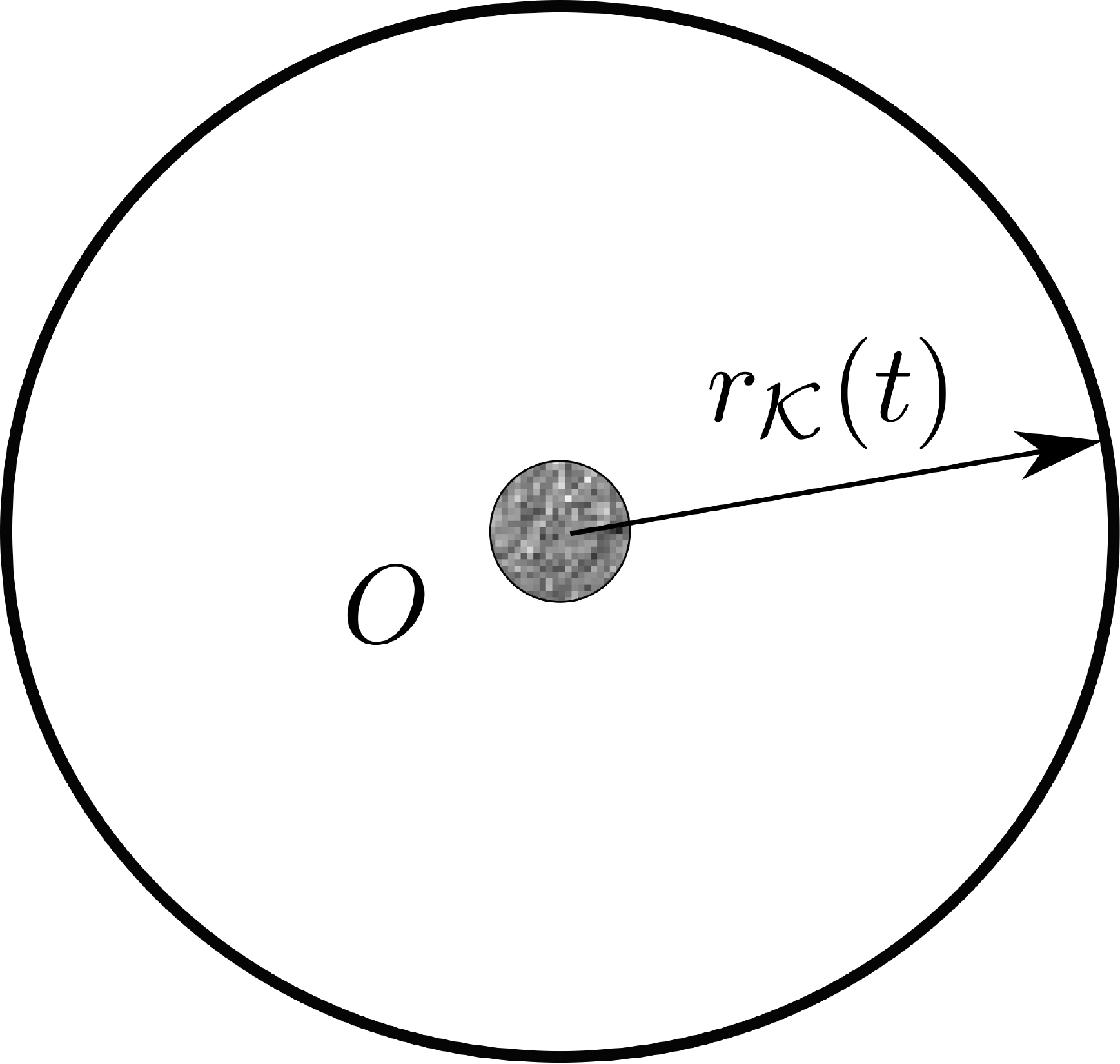} 
\caption{}\label{circular1}
   \end{subfigure}
    \begin{subfigure}[ht!]{\linewidth}
\def\svgwidth{\textwidth}
\centering
 \includegraphics[width=0.5\textwidth]{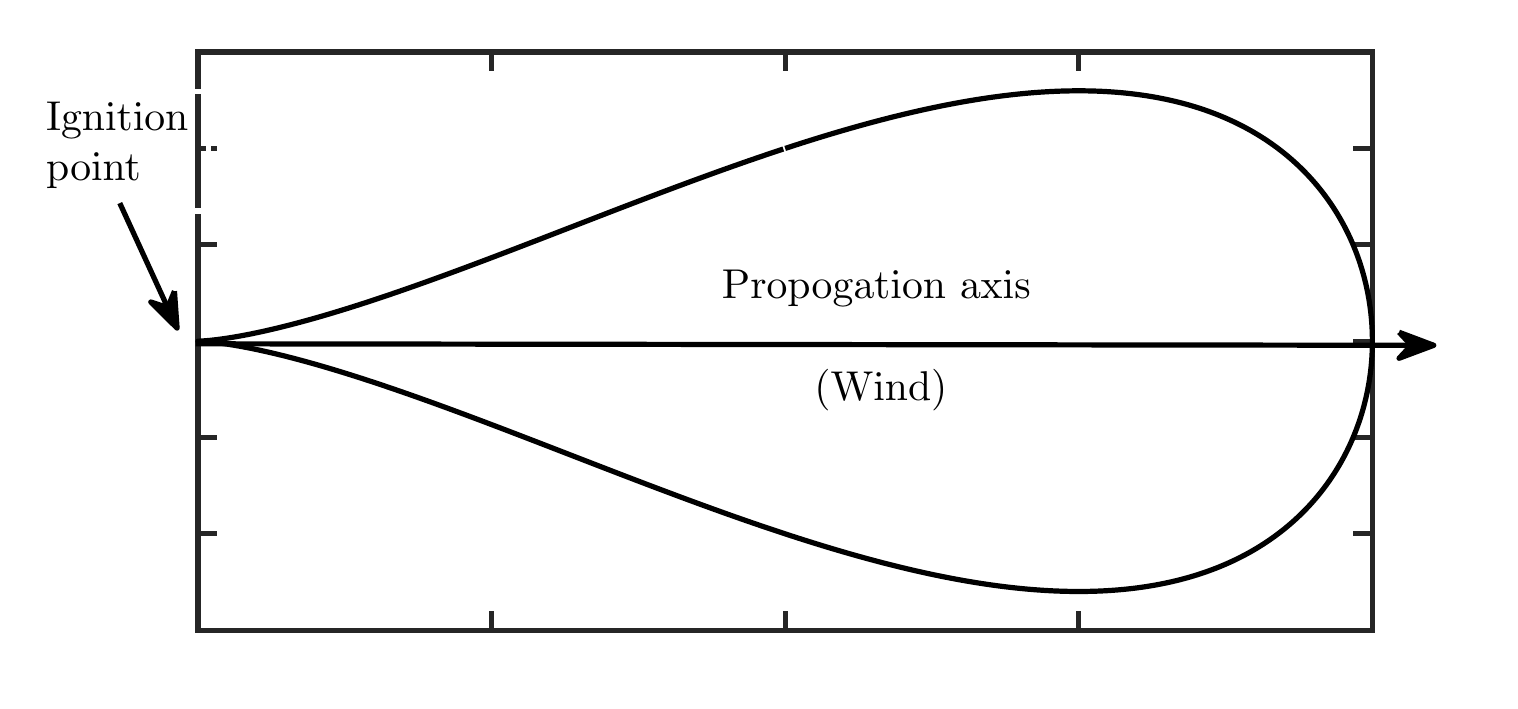}
 \caption{\small  \label{piriform20}}
\end{subfigure}
   \caption{\small Various propagation models of wildfire in a forest. (a)  In the presence of wind: elliptical model. (b) In the absence of wind: circular model (c) In the presence of dominant wind: piriform model}   
\end{figure}

The other simple fire propagation model is pear shaped or piriform propagation model (see Fig. \ref{piriform20}). It has been seen that if the air is dominating in a particular direction, the fire envelop attains a piriform shape \cite{ferguson2017}. Under the priform model, the fire envelop $\mathcal{K}(t)$ is given as:
\begin{align}
  \mathcal{K}(t)=\left\{ x,y:
    \begin{tabular}{l}
      $x=a(t)(1 + \sin \phi )$ \\
      $y=b(t)\cos \phi (1 + \sin \phi )$
\end{tabular},{0\leq\phi\leq2\pi}
\right\}.
\end{align}

Here, $a(t)$ and $b(t)$ are the two axes as given in \eqref{majminaxes}:

Area and perimeter of $\mathcal{K}(t)$ associated with the Periform model is given as:
\begin{align}
\area(\mathcal{K}(t))&=\pi a(t) b(t).\\
\ell(\mathcal{K}(t))&=\int\limits_0^{2\pi } {\sqrt {a^2(t)\cos^2 \theta + b^2(t){{(\cos 2\theta  -\cos \theta )}^2}} } \mathrm{d}\theta 
\end{align}

 The perimeter of the curve $\ell(\mathcal{K}(t))$ can be calculated by performing numerical integration.

\section{Coverage Analysis}

In this section, we will compute the sensing performance of the considered WFSN in terms of fire detection probability.  The fire detection probability $\fsprob$ of the system is defined as the probability that fire is detected by at least one sensor of the sensor network before the fire turns critical. Recall that an event of fire occurrence  is said to be not sensed at time $t$ if:
\begin{equation}
\xi \cap \mathcal{K}(t)= \phi.
\end{equation} 
Therefore, the fire detection probability is equal to the probability that any part of the fire region falls in the sensing region of at least one sensor at critical time $t_\crit$. 
Hence, the fire detection probability is given as
\begin{align}
\fsprob&=\mathbb{P}(\xi \cap \mathcal{K}(t_\crit)\ne \phi).
\end{align}

\subsection{Fire Sensing Probability at Time $t$}
Let us first compute the probability that a fire event is not sensed at time $t$ which is given by:
\begin{align}
\mathcal{G}(t)=&{}\mathbb{P}(\xi \cap \mathcal{K}(t)= \phi) \nonumber\\
=&\exp(-\lambda\mathbb{E}(\area(\hat{\mathsf{S}}\oplus\mathcal{K}(t)))\label{void}
\end{align}
where $\hat{\mathsf{S}}$ is the mirror image  of $\mathsf{S}$, $\oplus$ is the Minkowski addition \cite{HaenggiBook}.  Therefore, the probability of the set  $\mathcal{K}(t)$ being covered at time instant $t$ is:
\begin{align}
\fsprobt(t)=&{}1-\mathcal{G}(t).\nonumber \\
=&1-\exp(-\underbrace{\lambda\mathbb{E}[\area(\mathcal{K}(t)\oplus \hat{\mathsf{S}}])}_{\N(\mathcal{K}(t))}.\label{covered}
\end{align}
Note that $\N(\mathcal{K}(t))$ represents the mean number of sensors  that have detected the fire in their sensing range.  In  2-dimensional  case, the area of the Minkowski addition of the set $\mathcal{K}(t)$ with $\hat{S}=\mathcal{B}(0,r)$ can be evaluated by Steiner formula \cite{chiu2013stochastic}:
\begin{eqnarray}\label{Stiner}
\area(\mathcal{K}(t)\oplus \mathcal{B}(0,r))=\area(\mathcal{K}(t))+\ell(\mathcal{K}(t))r+\pi r^{2}.
\end{eqnarray}
Recall that $\ell(\mathcal{K}(t))$ is the boundary length of set $\mathcal{K}(t)$ and $\area(\mathcal{K}(t))$ is the area of $\mathcal{K}(t)$.
\subsection{Fire Detection Probability}
Now, the fire detection probability can be computed as
\begin{align}
\fsprob&=1-\exp(-{\lambda\mathbb{E}[\area(\mathcal{K}(t_\crit)\oplus \hat{\mathsf{S}}])}.\label{eq:fsprob}
\end{align}

\subsection{Critical Sensor Density}
 The critical sensor density ($\lambda
_{\mathrm{cr}}$) is defined as the density of sensors which can detect fire with probability $\tau$ before fire turns critical. 
The generalized expression for critical sensor density is given as follows:
\begin{align*}
\lambda_\crit: \tau=\fsprob&=1-\exp(-{\lambda_\crit\mathbb{E}[\area(\mathcal{K}(t_\crit)\oplus \hat{\mathsf{S}}])}.
\end{align*}
This gives
\begin{align}
\lambda_{\mathrm{cr}}(\tau)=\frac{1}{\mathbb{E}[\area(\mathcal{K}(t_{\mathrm{cr}})\oplus \hat{\mathsf{S}}]}\log\Big(\frac{1}{1-\tau}\Big).\label{eq:criticaldensity}
\end{align}

We will now analyze the specific fire propagation models proposed in Section III.

\subsection{Fire Detection Probability in the Absence of Wind (Circular Model)}
\begin{figure*}[ht!]
\centering
\begin{subfigure}[t]{.3\linewidth}
  \includegraphics[width=1\textwidth]{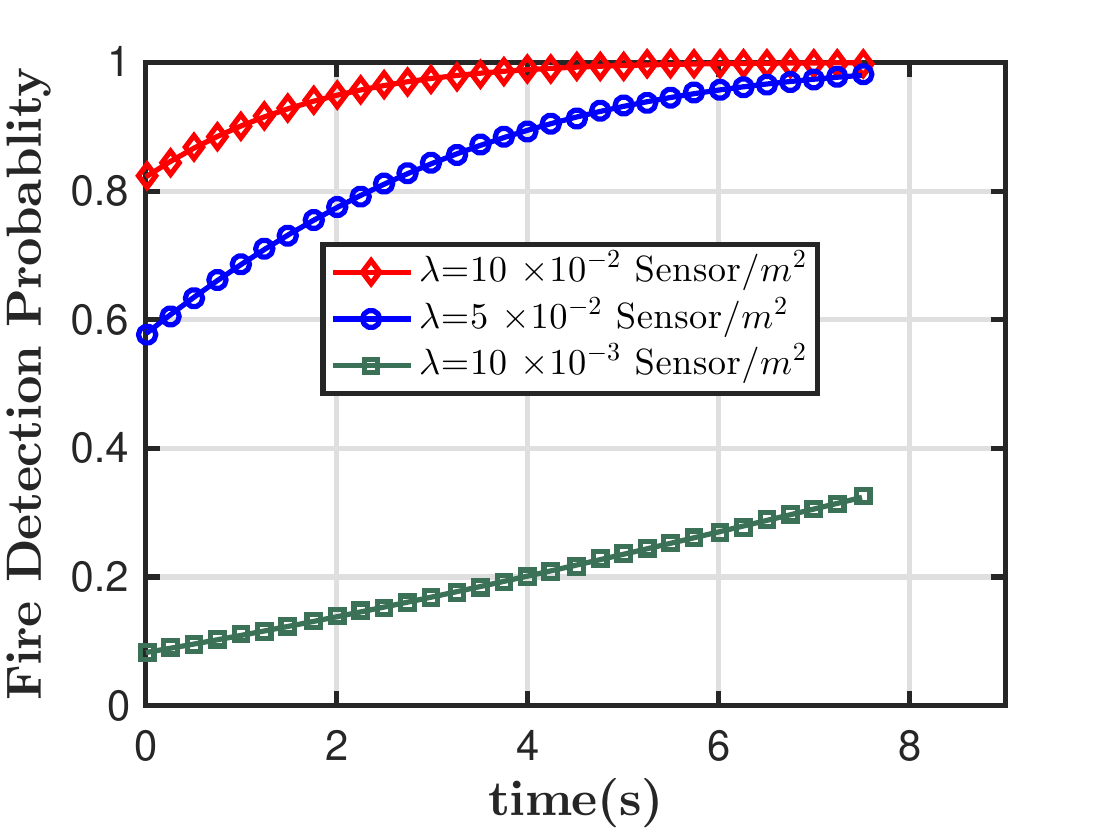}
  \caption{} \label{circularfirefront}
  \end{subfigure}
  \begin{subfigure}[t]{.3\linewidth}
  \includegraphics[width=1\textwidth]{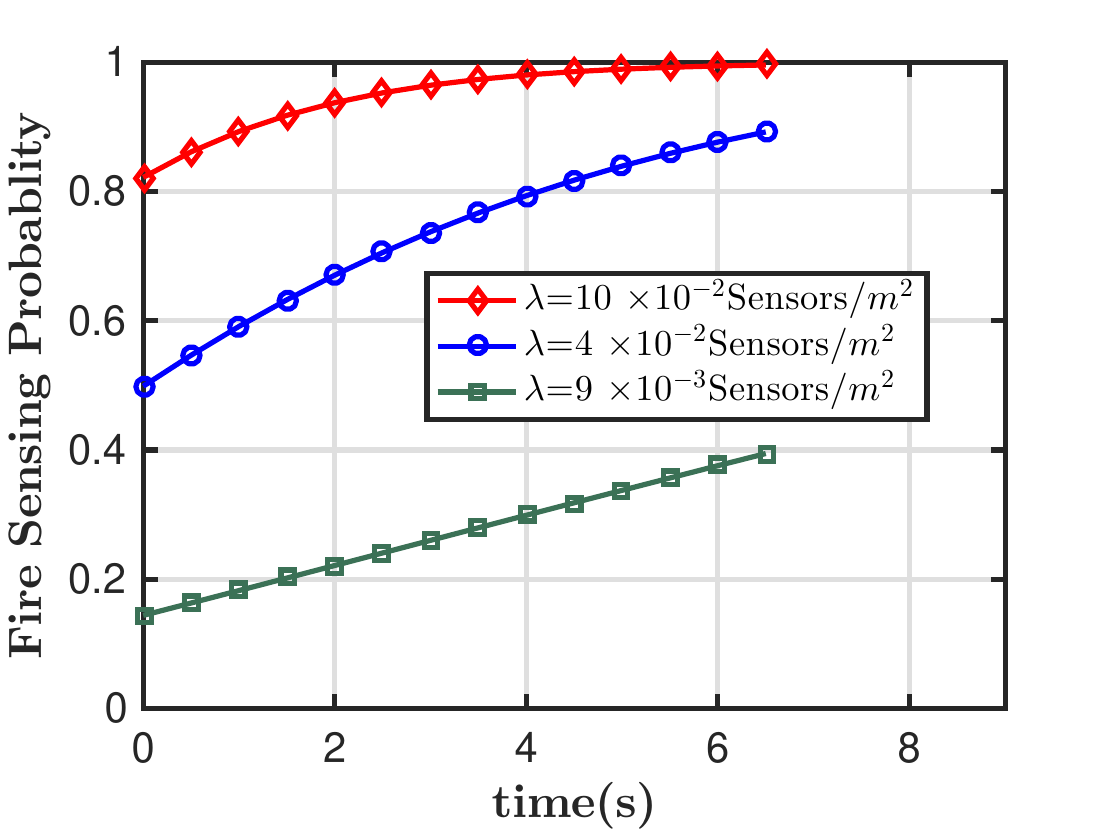}
  \caption{}\label{elliplsefiresensingprob}
   \end{subfigure}
 \begin{subfigure}[t]{.3\linewidth}
  \includegraphics[width=1\textwidth]{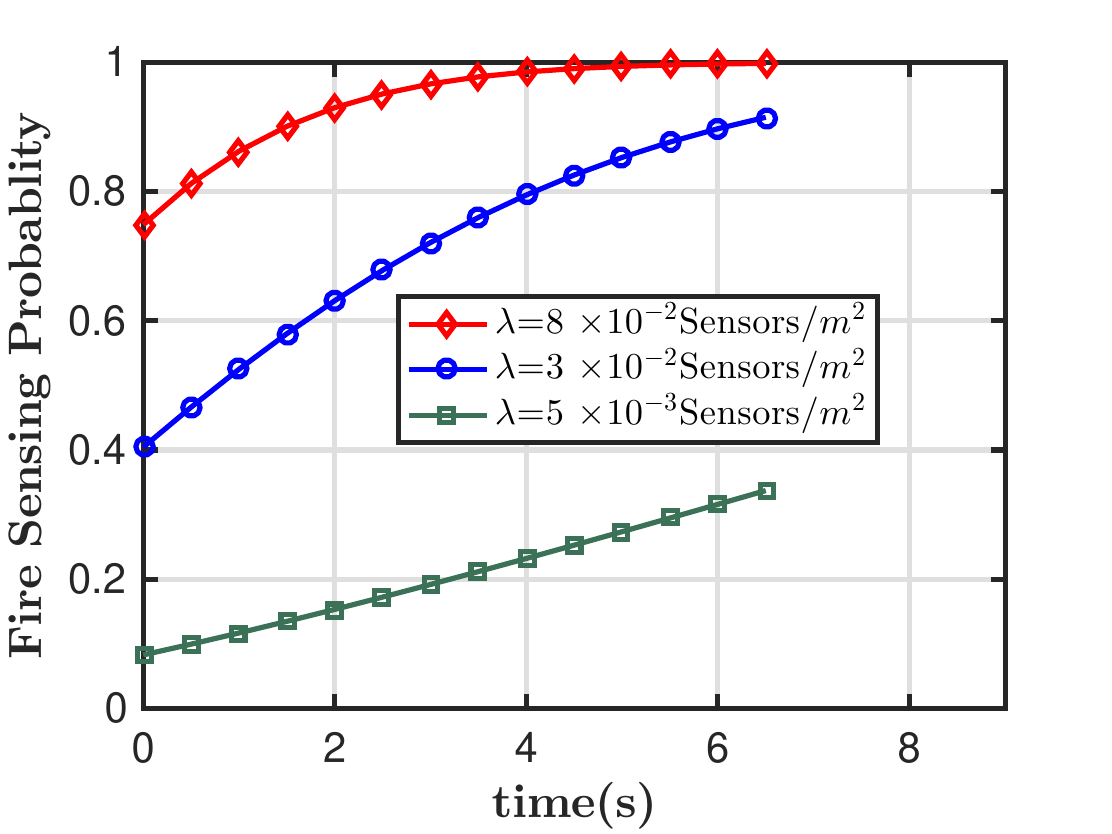}
  \caption{}\label{firesensensingprobpiriform}
   \end{subfigure}  
    \caption{\small Fire sensing probability $\fsprobt(t)$ with respect to time for various sensor density and fire propagation models. (a) Circular fire propagation model. WSN with $\lambda$= 10 $\times10^{-2}$ Sensors/$m ^{2}$  provides the fire detection probability $\fsprob$ close to 100 $\%$. (b) Elliptical fire propagation model. WSN with sensor density of 4 $\times10^{-2}$ Sensors/$m^{2}$ sensing probability is less than 60$\%$ but after 4-second sensing probability is above 80$\%$. (c) Piriform fire propagation model. For $\lambda$= 3 $\times10^{-2}$Sensors/$m^{2}$ which is less than the elliptical case initially the fire sensing probability is around 40$\%$ but within 4 seconds it reaches more than 80$\%$.} \label{fig:fig4}
\end{figure*}
Substituting area and perimeter for circular model in \eqref{Stiner}, the  mean number of sensors  detecting the fire can be computed as
\begin{align}
&\N(\mathcal{K}(t))={}\lambda\mathbb{E}[ \pi( \alpha t)^{2}+2 \pi \alpha t r+\pi r^{2}]\\
&={}\lambda\pi\left[ ( \alpha t)^{2}+2 \alpha t\mathbb{E}[ r]+ \mathbb{E}[ r^{2}]\right].
\end{align}
Using \eqref{eq:fsprob}, the fire detection probability is given as
\begin{align}
\fsprob(t)&=1-\exp\left(-\lambda\pi\left[ ( \alpha t_\crit)^{2}+2 \alpha t_\crit\mathbb{E}[ r]+ \mathbb{E}[ r^{2}]\right]\right).
\end{align}

The critical time is given as:
\begin{equation}
t_{\mathrm{cr}}\leq \frac{1}{\alpha}\sqrt{\frac{A_{\mathrm{cr}}}{\pi}}.
\end{equation}
Using the value of $t_\crit$ in \eqref{eq:criticaldensity}, critical sensor density $\lambda_\crit$ is given as
\begin{align}
\lambda_{\crit}(\tau)=&{}\frac{1}{\pi(\alpha t_\crit)^{2}+2 \pi \alpha t_\crit \mathbb{E}[r]+\pi \mathbb{E}[r^{2}]}\ln \Big(\frac{1}{1-\tau}\Big)\\
=& \frac{1}{A_{\mathrm{cr}}+2\sqrt{\pi A_{\mathrm{cr}}}\mathbb{E}[r]+\pi \mathbb{E}[r^{2}]}\ln \Big(\frac{1}{1-\tau}\Big).
\end{align}
Therefore, any $\lambda \geq \lambda_\crit$ will provide the fire detection  probability more than $\tau$.
\subsection{Fire Detection Probability in the Presence of Wind}
The mean number of sensor that can detect fire can be obtained
using \eqref{areaelliplse} and \eqref{perimerterellipse} and is given by as follows:
\begin{align} 
\N (\mathcal{K}(t)) &= \lambda\mathbb{E} [\pi a(t)b(t) + \pi r[3(a(t) + b(t))- \sqrt {(3a(t) + b(t))(a(t) + 3b(t)))} ] + \pi {r^2}]\nonumber\\
& =\lambda\pi\mathbb{E}\left[  {(\alpha t)^2}\left(1 + \frac{{{v_x}}}{V}\right) 
+  r\alpha t\left[3\left(2 + \frac{{{v_x}}}{V}\right)- \sqrt {(4 + \frac{{{v_x}}}{V})(4 + \frac{{3{v_x}}}{V})} \right]
 + {r^2}\right ]\nonumber\\
& =\lambda\pi\left[  {(\alpha t)^2}\left(1 + \frac{{{v_x}}}{V}\right) 
+  \mathbb{E}[r]\alpha t\left[3\left(2 + \frac{{{v_x}}}{V}\right)- \sqrt {(4 + \frac{{{v_x}}}{V})(4 + \frac{{3{v_x}}}{V})} \right]
 + \mathbb{E}[{r^2}]\right ]\label{sikforelliplse}.
\end{align}

The critical time $t_{\mathrm{cr}}$ is given as:
\begin{align}
t_{\mathrm{cr}}\leq &{} \frac{1}{\alpha}\sqrt{\frac{A_{\mathrm{cr}}}{\pi(1+\frac{v_{x}}{V})}}.
\label{eq:ellipsetcr}\end{align}
It is clear that critical time reduces in the presence of wind. Using \eqref{sikforelliplse} and \eqref{eq:ellipsetcr} in \eqref{eq:criticaldensity},   the critical sensor density can be computed as :

\begin{align}
{\lambda _{\mathrm{cr}}} &= \frac{1}{{[\pi a({t_{\mathrm{cr}}})b({t_{\mathrm{cr}}}) + \ell(\mathcal{K}({t_{\mathrm{cr}}}))\mathbb{E}[r] + \pi\mathbb{E}[ {r^2}]}}\log \left(\frac{1}{{1 - \tau }}\right)\nonumber\\
& = \frac{{\log (\frac{1}{{1 - \tau }})}}{\begin{array}{l}
{A_{\mathrm{cr}}} + \sqrt {\frac{{\pi {A_{\mathrm{cr}}}}}{{1 + \frac{{{v_x}}}{V}}}} \left[ 3(2 + \frac{{{v_x}}}{V}) - \sqrt {(4 + 3\frac{{{v_x}}}{V})(4 + \frac{{{v_x}}}{V})} \right] \mathbb{E}[r]+\mathbb{E}[{r^2}]
\end{array}}
\end{align}

\subsection{Fire Detection Probability for Piriform Model}

The critical time $t_{\mathrm{cr}}$ for piriform model is the same as the elliptical model. Now, the critical density is given as
\begin{equation}
\lambda_{\mathrm{cr}}=\frac{1}{A_{\mathrm{cr}}+\ell\left(\mathcal{K}\left( \frac{1}{\alpha}\sqrt{\frac{A_{\mathrm{cr}}}{\pi(1+\frac{v_{x}}{V})}}\right)\right)\mathbb{E}[r]+\pi \mathbb{E}[r^{2}]} \log\Big(\frac{1}{1-\tau}\Big).
\end{equation}
\section{Numerical Results}

In this section, we will evaluate fire detection probability and present some numerical results and insights for the models considered.  
The simulation parameters taken are listed in the Table \ref{General Parameter}.
 
\begin{table}[ht]\caption {\small Numerical Parameters} {\label{General Parameter}}
\begin{center}
\begin{tabular}{| l | l |}
 \hline
 \textbf{Parameter} & \textbf{Value}\\
 \hline
 Inner sensing range ($r_{\mathrm{in}}$)   & 2 meter  \\ \hline
 Outer sensing range ($r_{\mathrm{out}}$)&   4 meter  \\ \hline
 $\mathbb{E}[r]$ and $\mathbb{E}[r^2]$&  2.68 meter, 5.49 meter  \\ \hline
 Fire flame velocity ($\alpha$) & .33 meter/sec. \\ \hline
   Critical area ($A_\crit$)& 20 m$^{2}$ \\ \hline
   Wind velocity ($v_{x}$) in elliptical/piriform model & 3 m/s \\  \hline 
   Scaling factor (V) & 10 m/s    \\ \hline 
   
\end{tabular}

\end{center}
\end{table}

\textbf{Impact of sensor density:} 
Fig. \ref{fig:fig4} shows the variation of fire sensing probability $\fsprob(t)$ with time ($t$) for the different different scenarios: (a) in the absence of wind velocity (circular fire propagation), (b) in the presence of wind velocity (elliptical propagation) and (c) in the presence of dominant wind (piriform propagation).   It can be seen that increasing sensor density can significantly improve static detection probability which denotes the probability a fire is detected at its start only. For example, WSN with sensor density $\lambda$=.05 Sensors/$m^{2}$ can provide a static detection probability of $60\%$ in the absence of fire. It means that there is 60\% chance that fire start is immediately detected in the beginning.  After 3 second of fire event, the detection probability greater than 80$\%$. In the presence of wind, impact of increasing sensor density is less prominent. It can also be identified that having large sensor density does not have much influence on sensing probability on the other hand moderate sensor density have fairly good initial sensing probability and rapidly increases with time.

\begin{figure}[ht!]
  	\centering
  \includegraphics[width=.5\textwidth]{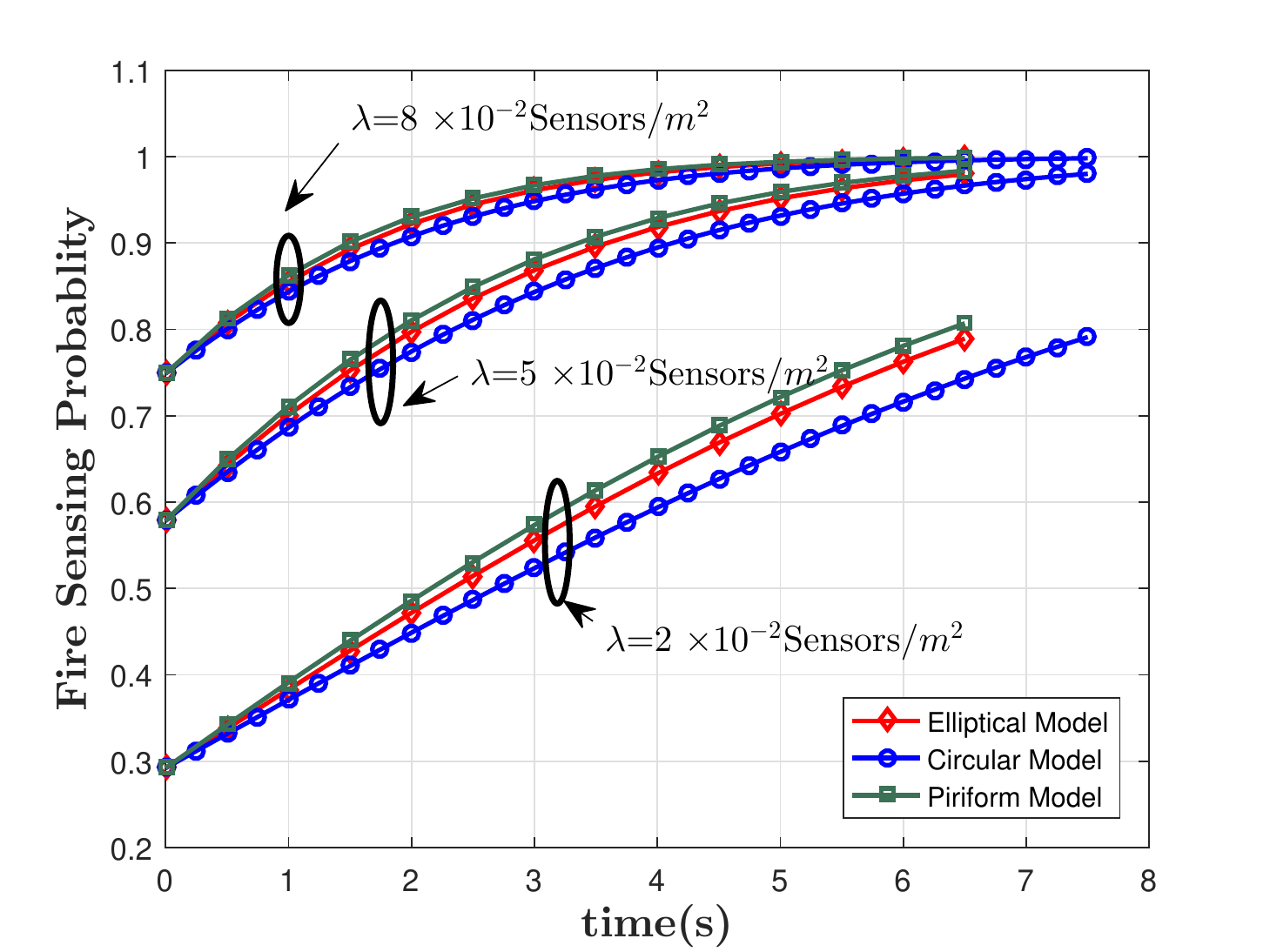}
   \caption{\small Comparative analysis between circular, elliptical and piriform models.   In the presence of the wind, the sensing probability is higher than the no-wind case.}   \label{comparaitivegraph}
\end{figure}

\textbf{Comparison of three scenarios:}
Fig. \ref{comparaitivegraph} shows the comparison of three scenarios. 
 In the absence of any wind, the critical time ($t_{\mathrm{cr}}$) is 7.6 s.   The critical time in the presence of wind is 6.7 second which is less than as compared to the no-wind case. It is due to the faster spread of fire due to wind giving even smaller window to detect fire. 
 The critical time ($t_{\mathrm{cr}}$) for piriform type propagation is the same as the elliptical  propagation. However, piriform type propagation gives better coverage due to larger perimeter-to-area ratio making it easier for sensors to detect this fire in the same time.

\begin{figure}[ht!]
	\centering
  \includegraphics[width=.5\textwidth]{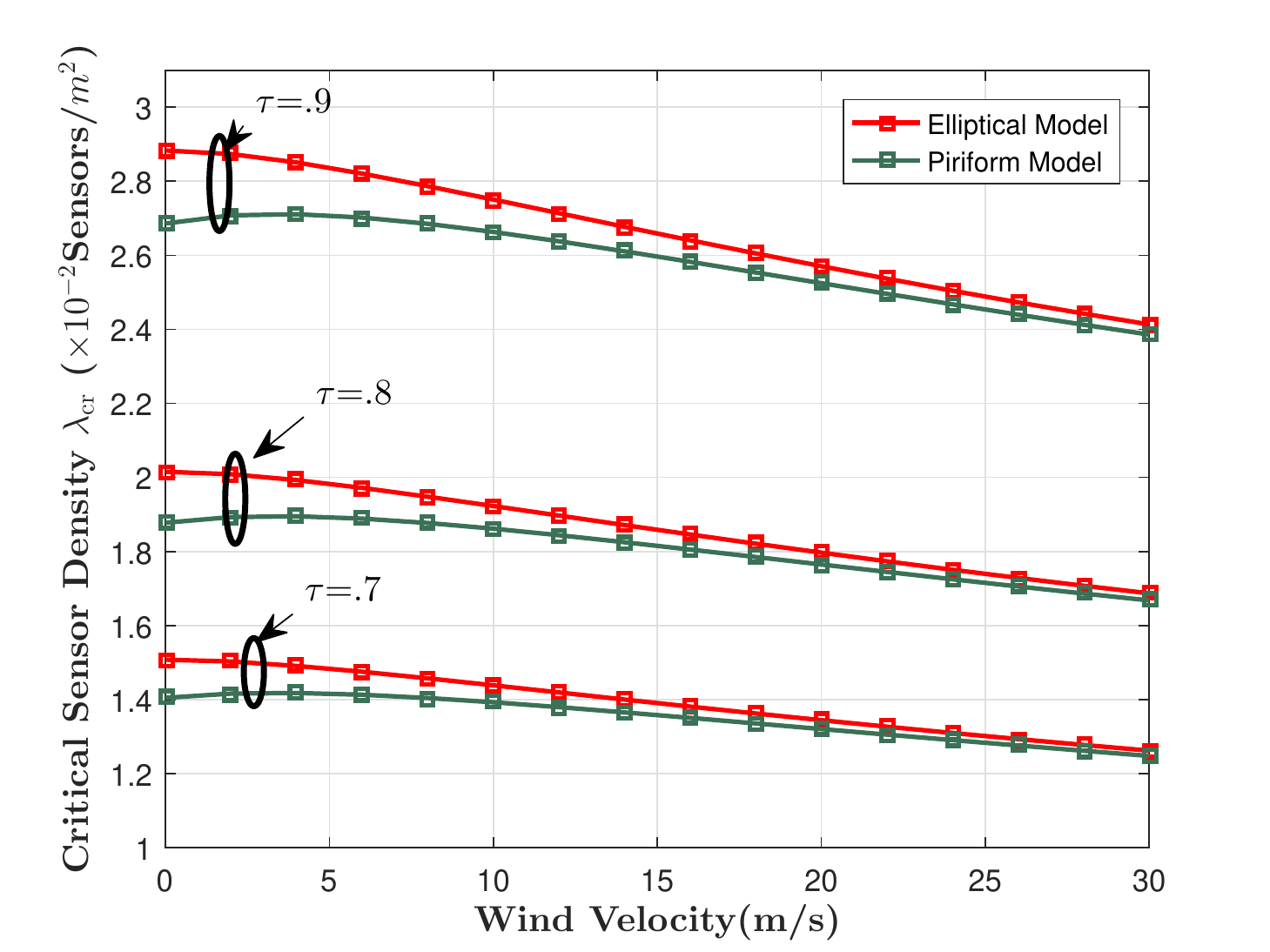} 
   \caption{\small Comparative analysis between elliptical and piriform model: Critical sensor density with respect to wind velocity for different values of fire detection probablity thresold ($\tau$).  In high wind areas, less density of sensors are required to achieve similar fire detection probability thresold.}   \label{criticalsensorcompr}
\end{figure}

\textbf{Impact of wind velocity on critical sensor density:}
Fig.  \ref{criticalsensorcompr} shows the impact of wind velocity on critical sensor density for various propagation models. Recall that critical sensor density corresponding to the zero wind velocity refers to the circular propagation. 
 The critical sensor density reduces in piriform type propagation as compared to elliptical type propagation of fire.  This is consistent with the previous result. The another important observation is due to the impact of wind on the propagation of fire. It seems that in high winds, critical time reduces which is one of critical concern. However, due to high rate of fire spread, the fire detection probability threshold also increases. It can be seen that the wind velocity can effectively help in the detection of wildfire. In the case of piriform type propagation, a non-monotonic behavior of critical density with respect to wind velocity can be observed.

\section{Conclusions}\label{section5}

In this paper, we have considered a WSN with fire sensors for early detection of the forest fire. We present an analytical framework based on the Boolean-Poisson model, with the elliptical, circular and piriform fire flame propagation. Using the framework, we compute the critical sensor density which needs to be deployed in the forest to ensure a certain minimum fire detection probability. It has identified that in the presence of wind, critical time $t_{\mathrm{cr}}$ to detect fire decreases but the fire sensing probability also increases in comparison to the case without the wind.

\bibliographystyle{ieeetran}

\vspace{12pt}
\color{red}

\end{document}